\documentclass[aps,prd,superscriptaddress,twoside,twocolumn,nofootinbib,
]{revtex4-1}

\usepackage{dcolumn}
\usepackage{ulem} 
\usepackage{float}
\usepackage[usenames]{color}
\usepackage{amsmath}
\usepackage{graphicx}
\allowdisplaybreaks

\begin{document}

\title{ 
Light Quarkonium - Glueball Mixing from a Holographic QCD
} 

\author{Takashi Yamaguchi}
\email{yamaguchi@hken.phys.nagoya-u.ac.jp}
\affiliation{Department of Physics,  Nagoya University, Nagoya, 464-8602, Japan}

\author{Shinya Matsuzaki}
\email{synya@hken.phys.nagoya-u.ac.jp}
\affiliation{Institute for Advanced Research, Nagoya University, Nagoya 464-8602, Japan.}
\affiliation{Department of Physics,  Nagoya University, Nagoya, 464-8602, Japan}

\date{\today}

\begin{abstract}
We study the mixing structure of isospin-singlet scalars, the light quarkonium $(\bar{q}q)$
and glueball $(gg)$ in two-flavor QCD, based on a holographic model of bottom-up hard-wall type. 
In the model the pure quarkonium and glueball states are unambiguously defined in terms of the different $U(1)_A$ charges 
in the restoration limit of the chiral $U(2)_L \times U(2)_R$ symmetry, in which the quarkonium gets massless as the chiral partner of the pion. 
Hence the $\bar{q}q$-$gg$ mixing arises in the presence of the nonzero chiral condensate or pion decay constant.  
At the realistic point where the pion decay constant and other hadron masses reach the observed amount, 
we predict the tiny mixing between the lightest quarkonia and glueball: 
The smallness of the mixing is understood by the slightly small ratio of the chiral and gluon condensate scales.   
The low-lying two scalar masses are calculated to be $\simeq 1.25$ GeV and $\simeq 1.77$ GeV, 
which are compared with masses of $f_0(1370)$ and $f_0(1710)$. 
Our result implies that $f_0(1710)$ predominantly consists of glueball.          
\end{abstract}
\maketitle

\section{Introduction}

The spectrum structure of the low-lying scalars around the mass 1 GeV  
still remains unsolved in the low-energy QCD, and has  
currently been pursued extensively by several approaches. 
Of particular interest are  the isospin-singlet scalars, denoted as $f_0$'s  
in the quark model~\cite{Eidelman:2004wy}, 
since they can arise as mixtures having the same quantum number $J^{PC}=0^{++}$,    
such as the light two-flavor $(u,d)$ quarkonium ($\bar{q}q$), 
glueball ($gg$).  
Understanding such a rich low-lying isoscalar spectrum is therefore 
of great importance, which is tied to revealing some part of the dynamics of low-energy QCD.

 The straightforward investigation for the mixing structure  
has so far been performed on the full lattice QCD simulations~\cite{Hart:2006ps}, 
which focuses on the mixing $\bar{q}q$ and $gg$ states, 
and also has been made by using QCD sum rules (Ref.~\cite{Ochs:2013gi} for a recent review). 
The quantitative estimate on the mixing angle is, however, 
still less accurate so that one cannot say anything about the constituent structure.   
On the one hand, other recent approaches  
based on phenomenological hadron models 
have attempted to give some insight on the $\bar{q}q$-$gg$ mixing.  
By performing global fit to model parameters 
with use of the phenomenological inputs such as currently observed scalar-decay properties, 
it has been indicated that  
$f_0(1710)$ is almost constructed from the pure glueball state~\cite{Janowski:2014ppa}. 
Still, however, the issue is controversial so it needs more indication from some different approaches.

In this paper, we study the mixing between the quarkonium and glueball states 
based on the gauge-gravity duality~\cite{Maldacena:1997re,Witten:1998qj}, 
so-called holographic QCD.    
We employ a holographic model proposed in Ref.~\cite{Haba:2010hu}, 
an improved version of 
bottom-up hard-wall type proposed in Refs.~\cite{Erlich:2005qh,Da Rold:2005zs}. 
The model has succeeded in 
reproducing the characteristic features of QCD, such as the observed meson interaction properties as well as 
the ultraviolet-asymptotic behavior of QCD Green functions. 
What should be remarked in the model of~\cite{Haba:2010hu} is, in particular, simultaneous incorporation of the five-dimensional bulk fields dual to  
the gluon and the chiral condensate operators, $G_{\mu\nu}^2$ and $\bar{q}q$. 
As emphasized in~\cite{Haba:2010hu},  
one can thereby reproduce the ultraviolet scaling of QCD current correlators including terms along with the condensate of $G_{\mu\nu}^2$
as well as the leading logarithmic term and the chiral condensate term. 
The isospin-singlet scalars, 
arising as fluctuation modes around the condensates in the holographic bulk, thus include both the quarkonium and glueball states, 
hence the mixing among them can be evaluated straightforwardly.

We analyze the two-flavor QCD case, in which the chiral $U(2)_L \times U(2)_R$ symmetry is spontaneously broken down to 
the vectorial one $U(2)_V$ via the nonzero vacuum expectation value of a bulk-scalar dual to the $\bar{q}q$, accordingly 
due to the nonzero pion decay constant $f_\pi$. 
In the model the pure quarkonium and glueball states are unambiguously defined in terms of the different $U(1)_A$ charges 
in the restoration limit of the chiral $U(2)_L \times U(2)_R$ symmetry ($\langle \bar{q}q \rangle$ and  $f_\pi \to 0$), 
in which the quarkonium gets massless, reflecting  the chiral partner of the pion. 
Once the chiral condensate develops from zero, the $\bar{q}q$-$gg$ mixing is turned on and grows monotonically with the chiral condensate.   
At the realistic point where $f_\pi$ and masses of other mesons 
such as the vector/axialvector mesons reach the desired amount,   
we find the tiny mixing between the lightest quarkonium and glueball,   
which is consistent with the large $N_c$ picture on that of basis the holographic model has been established. 
 This is the definite prediction obtained     
without any phenomenological inputs such as observed isospin-singlet scalar decay properties, in contrast to 
other effective-hadron model approaches~\cite{Janowski:2011gt,Janowski:2012yg,Parganlija:2012fy,Janowski:2014ppa}.  
We further provide a new insight on the small mixing in terms of generic quantities in QCD:  
the smallness of the mixing is understood by the somewhat small ratio of the chiral 
and gluon condensate scales at the realistic point.

At the realistic point the lowest-lying two scalar masses are calculated to be $\simeq 1.25$ GeV and $\simeq 1.77$ GeV, 
which are compared with masses of $f_0(1370)$ and $f_0(1710)$.    
The lightest isospin-singlet scalar 
is almost completely degenerate with the isospin-triplet scalar, $a_0$ meson, which is due to the extremely-small mass split induced from 
the vanishingly small mixing among the isospin-singlet scalars. 
 These two lower masses $\simeq 1.2$ GeV can actually be lifted up to the desired value around $\simeq 1.3 - 1.4$ GeV, when
 the mixing with a four-quark state would be taken into account~\cite{Black:1999yz}.

Our result thus implies that $f_0(1710)$ predominantly consists of glueball with the mass around $\simeq 1.7 - 1.8$ GeV, 
in accord with the pure glueball mass estimate by lattice simulations~\cite{Morningstar:1999rf}, 
a recent study based on a holographic QCD of top-down type~\cite{Brunner:2015oqa}  
and  a different approach based on a phenomenological model~\cite{Janowski:2014ppa}.

This paper is organized as follows: 
 In Sec.~\ref{Model} we give a brief review of the holographic model in Ref.~\cite{Haba:2010hu} and summarize 
 things necessary for analysis of the isospin-singlet scalar mixing. 
 In Sec.~\ref{The-induced-effective-Lagrangian}  
we define the pure quarkonium and glueball states and introduce effective interactions describing the mixing of them  
deduced from the present holographic model. 
The effective mass matrix, obtained by keeping a few low-lying scalars, is then evaluated 
in Sec.~\ref{Analysis-of-the-mixing-structure}.  
Summary of this paper is given in Sec.~\ref{Summary}.

\section{Model: Preliminaries}  
\label{Model}

We begin by reviewing the holographic model proposed in Ref.~\cite{Haba:2010hu} 
and list some equations and formulas necessary for the later discussions.

The model in Ref.~\cite{Haba:2010hu} is 
based on deformations of 
a bottom-up approach for successful holographic dual 
of QCD~\cite{Erlich:2005qh,Da Rold:2005zs}. 
The model we shall employ is described as $U(2)_L \times U(2)_R$ gauge theory 
which is defined on the five-dimensional anti-de-Sitter (AdS) space-time. 
The five-dimensional space-time is characterized by 
the metric $ds^2= g_{MN} dx^M dx^N 
= \left(L/z \right)^2\big(\eta_{\mu\nu}dx^\mu dx^\nu-dz^2\big)$ 
with $\eta_{\mu\nu}={\rm diag}[1, -1, -1,-1]$. 
Here,  $M$ and $N$ ($\mu$ and $\nu$) represent five-dimensional (four-dimensional) 
Lorentz indices, and $L$ denotes the curvature radius of the AdS background.
The fifth direction, denoted as $z$, is compactified on an interval extended 
from the ultraviolet (UV) brane located at $z=\epsilon$ 
to the infrared (IR) brane at $z=z_m$, i.e., $ \epsilon \leq z \leq z_m  $. 
The UV cutoff $\epsilon$ will be taken to be 0 after all calculations are done.

  Besides the bulk left- ($L_M$) and right- ($R_M$) gauge fields, 
we introduce two bulk scalars $\Phi_{qq}$ and $\Phi_{gg}$. 
The $\Phi_{qq}$ transforms as a bifundamental 
representation field under the $U(2)_L \times U(2)_R$ gauge symmetry, 
and therefore is dual to the quark bilinear operator $\bar{q} q$ having 
the $U(1)_A$ charge equal to 2. 
The $\Phi_{gg}$ is, on the other hand, chiral and $U(1)_A$-singlet, dual to 
the gluon condensate operator $G_{\mu\nu}^2$. 
The mass-parameter for these two bulk scalars $M_{\Phi_{qq}}$ and $M_{\Phi_{gg}}$ 
are then holographically given as  
$M_{\Phi_{qq}}^2=- 3/L^2$ and  
$M_{\Phi_{gg}}^2 = 0,$ 
which reflect the scaling dimensions for $\bar{q}q$ and $G_{\mu\nu}^2$, respectively.

The action of the model is thus written as~\cite{Haba:2010hu} 
\begin{equation} 
  S_5 = S_{\rm bulk} + S_{\rm UV} + S_{\rm IR} 
  \,, \label{S5}
\end{equation}
where 
\begin{eqnarray} 
  S_{\rm bulk} 
  &=& 
  \int d^4 x \int_\epsilon^{z_m} dz 
  \sqrt{g} 
  \frac{1}{g_5^2} \, e^{c g_5^2 \Phi_{gg}} 
 \Bigg[ 
\frac{1}{2} \partial_M \Phi_{gg} \partial^M \Phi_{gg} 
\nonumber \\ 
&& 
+ {\rm Tr}[D_M \Phi_{qq}^\dag D^M \Phi_{qq} - M_{\Phi_{qq}}^2 \Phi_{qq}^\dag \Phi_{qq} ] 
\nonumber \\ 
&&
  - \frac{1}{4} {\rm Tr}[L_{MN}L^{MN} + R_{MN}R^{MN}] 
 \Bigg] 
 \,,  \label{S:bulk} \\ 
 S_{\rm IR} 
&=&  
   \int d^4 x \int_\epsilon^{z_m} dz  \, \delta (z -z_m) 
  \sqrt{-\tilde{g}} \, {\cal L}_{\rm IR} 
\,, \label{S:UVIR}
\end{eqnarray} 
 with the boundary-induced metric $\tilde{g}_{\mu\nu}=(L/z)^2 \eta_{\mu\nu}$.  
The covariant derivative acting on $\Phi_{qq}$ in Eq.(\ref{S:bulk}) 
 is defined as $D_M\Phi_{qq}=\partial_M \Phi_{qq}+iL_M\Phi_{qq}-i\Phi_{qq} R_M$, where 
$L_M(R_M)\equiv L_M^a(R_M^a) T^a$ with $T^a=\sigma^a/\sqrt{2}$ $(a=0,1,2,3)$ being the generators of 
$SU(2)$ and $\sigma^0 =1_{2 \times 2}$ normalized as 
${\rm Tr}[T^a T^b]=\delta^{ab}$. 
$L(R)_{MN}$ is the five-dimensional field strength which is defined as 
$L(R)_{MN} = \partial_M L(R)_N - \partial_N L(R)_M 
 - i [ L(R)_M, L(R)_N ]$, and 
$g$ is defined as $g={\rm det}[g_{MN}]= (L/z)^{10}$. 
The UV boundary action $S_{\rm UV}$ in Eq.(\ref{S5}) plays a role of 
the UV regulator to absorb the UV-divergent $\epsilon$ terms 
arising from the five-dimensional bulk dynamics, which we will not specify. 
The IR boundary action $S_{\rm IR}$ is introduced so as to 
realize minimization of the bulk potential by 
nonzero chiral condensate~\cite{Haba:2010hu} with the IR Lagrangian: 
\begin{eqnarray}  
  {\cal L}_{\rm IR} 
  &=& - e^{c g_5^2 \langle \Phi_{gg} \rangle} \left( - m_b^2 {\rm Tr}[|\Phi_{qq}|^2] + \lambda {\rm Tr}[|\Phi_{qq}|^2]^2 
\right)
\,. \label{L:IR}
\end{eqnarray} 
The gauge coupling $g_5$ and a parameter $c$ appearing in the action 
are fixed by matching with the UV asymptotic behavior of vector/axialvector current correlators
as~\cite{Haba:2010hu}  
\begin{equation}
\frac{L}{g_5^2} = \frac{N_{c}}{12\pi^2},\quad c = -\frac{L}{16\pi g_5^2} = - \frac{N_c}{192 \pi^3}\,,
\label{matching}
\end{equation}
where the latter has been determined from terms damping along with the gluon condensate.

\subsection{Vacuum expectation values of scalars}

The bulk scalar fields $\Phi_{qq}$ and $\Phi_{gg}$  
are parametrized as   
\begin{eqnarray}
\Phi_{qq}(x,z) 
&=& \frac{1}{\sqrt{2}} \left(v_{qq}(z)+ \sigma_{qq}(x,z) + a_{0}^i(x,z) \sigma^i \right) 
\nonumber \\ 
&& \times e^{i \pi^a(x,z) \sigma^a /v_{qq}(z)}, 
\qquad (i=1,2,3) 
\label{KK:qq}
\end{eqnarray} 
\begin{equation} 
\chi (x,z) \equiv e^{\frac{c g_5^2}{2} \Phi_{gg} } =v_{\chi}(z) \, e^{
\sigma_{gg}(x,z)}, 
\label{KK:gg}
\end{equation}  
with the vacuum expectation values, $v_{qq}=\sqrt{2} \langle \Phi_{qq} \rangle$ and 
$v_{\chi} = \langle \chi \rangle = e^{\frac{c g_5^2}{2} \langle \Phi_{gg} \rangle}$.  
We hereafter disregard pion fields $\pi^a$ which will not be relevant for the present study. 
Plugging these expansion forms into the bulk action $S_{\rm bulk}$ Eq.(\ref{S:bulk}) we find 
the coupled equations of motion for the vacuum expectation values $v_{qq}$ and $v_{\chi}$, 
\begin{eqnarray} 
&& v_{\chi}^{-2} \partial_z \left( \frac{1}{z^3} v_{\chi}^2 \partial_z v_{qq}  \right) + \frac{3}{z^5} v_{qq} = 0 
\,, \nonumber \\ 
&& 
v_\chi^{-1} \partial_z \left( \frac{1}{z^3} \partial_z v_{\chi} \right) + 2 s^2 \, L^2\, \left( \frac{(\partial_z v_{qq})^2}{z^3} 
+ \frac{3 v_{qq}^2}{z^5}  \right) 
= 0 
\,, \label{EOM:v}  
\end{eqnarray}
where 
\begin{equation} 
 s \equiv \frac{cg_5^2}{2 L} = - \frac{1}{32 \pi} \simeq - 0.01 
 \,. \label{s:def}
\end{equation}
The last equality of the second line in Eq.(\ref{s:def}) follows from Eq.(\ref{matching}).    
The boundary conditions for $v_{qq}$ and $v_{\chi}$ are chosen 
(in the limit where the sources for $\bar{qq}$ and $G_{\mu\nu}^2$ are turned off) 
as~\cite{Haba:2010hu} 
\begin{eqnarray} 
 v_{qq}(\epsilon) &=& 0 \,, \qquad v_{qq}(z_m) = \frac{\xi}{L} 
 \,, \nonumber \\ 
 v_{\chi}(\epsilon) &=& 1 \,, \qquad v_{\chi}(z_m) = 1 + G 
 \,, \label{BC:v}
\end{eqnarray} 
with the IR boundary values $\xi$ and $G$ which are holographically related to 
the chiral and gluon condensates as shown in Ref.~\cite{Haba:2010hu}. 
 Looking at the numerical value of $s$ in Eq.(\ref{s:def}), as done in Ref.~\cite{Haba:2010hu} 
one may neglect the second term of ${\cal O}(s^2)={\cal O}(10^{-4})$ 
 in the equation of motion for $v_{\chi}$ in Eq.(\ref{EOM:v}) to decouple the $v_{qq}$ and $v_{\chi}$. 
 Then the approximate equations of motion can be analytically solved to be~\cite{Haba:2010hu} 
\begin{eqnarray} 
 v_{qq} &\simeq& \frac{ \xi}{L} (1+G) \cdot \frac{(z/z_m)^3}{1 + G (z/z_m)^4} 
 \,, \nonumber \\  
 v_{\chi} & \simeq & 1 + G (z/z_m)^4 
 \,, \label{v:approx}
\end{eqnarray} 
with the boundary conditions in Eq.(\ref{BC:v}) incorporated. 
[We have checked that, by numerically solving the coupled equation in Eq.(\ref{EOM:v}) keeping the $s^2$ term,
this approximation is satisfied within a few percent level 
in a wide range of the parameter space we will consider later. ]

As noted above, the nonzero value of the chiral condensate parameter $\xi$ can be realized by adjusting 
the IR potential in $S_{\rm IR}$ of Eq.(\ref{S5}): putting the solutions in Eq.(\ref{v:approx}) into the action $S_{5}$ in Eq.(\ref{S5}) 
and performing the minimization of the potential energy, one finds the stationary condition~\cite{Haba:2010hu} 
\begin{equation} 
 \xi^2 = \frac{1}{\lambda} \left[ 
 m_b^2 L - \frac{N_c}{12 \pi^2} \left( 3- \frac{4G}{1+G}  \right)
\right]\,. 
\label{stationary} 
\end{equation}

\subsection{Scalar sector}

The five-dimensional bulk scalar fields $\sigma_{qq}(x,z)$, 
$\sigma_{gg}(x,z)$ and $a_{0}^i(x,z)$ in Eqs.(\ref{KK:qq}) and (\ref{KK:gg}) 
can be expanded in terms of normalizable modes, corresponding to QCD scalar mesons:  
\begin{eqnarray} 
  \sigma_{qq}(x,z) &=& \sum_{n} \sigma_{qq}^{(n)}(x) f^{(n)}_{\sigma_{qq}}(z) 
\,,  \nonumber \\ 
    \sigma_{gg}(x,z) &=& \sum_{n} \sigma_{gg}^{(n)}(x) f^{(n)}_{\sigma_{gg}}(z) 
\,, \nonumber \\ 
  a_{0}^i(x,z) &=& \sum_{n} a_{0}^{i(n)}(x) f^{(n)}_{a_{0}}(z) 
\,, \label{KK:decomp}
\end{eqnarray}
with the wavefunctions $f^{(n)}_{\sigma_{qq}}, f^{(n)}_{\sigma_{gg}}, f^{(n)}_{a_{0}}$ having 
the normalizable UV boundary conditions,  
\begin{equation}
f^{(n)}_{\sigma_{qq}}(\epsilon) = 0\,, \qquad 
f^{(n)}_{a_{0}}(\epsilon) = 0\,, \qquad 
f^{(n)}_{\sigma_{gg}}(\epsilon) = 0 
\,. \label{UVBC:f}
\end{equation} 
Using Eq.(\ref{KK:decomp}) we expand the bulk scalar sector of the action $S_{\rm bulk}$ in Eq.(\ref{S5}) to find the induced effective Lagrangian 
in four-dimension:  
\begin{eqnarray} 
 {\cal L}_{\rm scalar} 
&=& {\cal L}_{\sigma_{qq}^2} + {\cal L}_{a_{0}^2} + {\cal L}_{\sigma_{gg}^2} + {\cal L}_{\sigma_{qq}\sigma_{gg}} 
 \,, \label{L:scalar}
\end{eqnarray} 
where 
\begin{eqnarray}  
{\cal L}_{ \sigma_{qq}^2} &=&   
\sum_{n} \frac{1}{2} A_{\sigma_{qq}}^n \left( \partial_\mu \sigma_{qq}^{(n)} \right)^2 - \frac{1}{2} B_{\sigma_{qq}}^n (\sigma_{qq}^{(n)})^2 
\,, \nonumber \\ 
{\cal L}_{ a_{0}^2} &=&   
\sum_{n} \frac{1}{2} A_{a_{0}}^n \left( \partial_\mu a_{0}^{i(n)} \right)^2 - \frac{1}{2} B_{a_{0}}^n (a_{0}^{i(n)})^2 
\,, \nonumber \\ 
{\cal L}_{ \sigma_{gg}^2} &=&   
\sum_{n} \frac{1}{2} A_{\sigma_{gg}}^n \left( \partial_\mu \sigma_{gg}^{(n)} \right)^2 - \frac{1}{2} B_{\sigma_{gg}}^n (\sigma_{gg}^{(n)})^2 
\,, \nonumber \\ 
{\cal L}_{ \sigma_{qq} \sigma_{gg}} &=&   
\sum_{m, n} C_{\sigma_{qq}\sigma_{gg}}^{(m, n)} \sigma_{qq}^{(m)} \sigma_{gg}^{(n)} 
\,, \label{Lags:scalar}
\end{eqnarray} 
with 
\begin{eqnarray} 
A_{\sigma_{qq}}^n &=& 
\frac{N_c}{12\pi^2}  \int  \frac{d z}{z} 2 \left( \frac{L}{z} \right)^2 v_\chi^2 (f^{(n)}_{\sigma_{qq}})^2 
\,, \nonumber \\ 
A_{a_{0}}^n &=& 
\frac{N_c}{12\pi^2}  \int  \frac{d z}{z} 2 \left( \frac{L}{z} \right)^2 v_\chi^2 (f^{(n)}_{a_{0}})^2 
\,, \nonumber \\ 
A_{\sigma_{gg}}^n &=& 
\frac{N_c}{12\pi^2}  \cdot 
\frac{1}{s^2} \cdot 
 \int  \frac{d z}{z}  \left( \frac{L}{z} \right)^2 v_\chi^2 (f^{(n)}_{\sigma_{gg}})^2 
\,, \nonumber \\ 
B_{\sigma_{qq}}^n &=& 
\frac{N_c}{12\pi^2}  \int \frac{d z}{z} 2 \left( \frac{L}{z} \right)^2 v_\chi^2 
\nonumber \\ 
&& \times 
\left[ 
( \partial_z f^{(n)}_{\sigma_{qq}})^2 - \frac{3}{z^2} (f^{(n)}_{\sigma_{qq}})^2 \right]
\,, \nonumber \\ 
B_{a_{0}}^n &=& 
\frac{N_c}{12\pi^2}  \int \frac{d z}{z} 2 \left( \frac{L}{z} \right)^2 v_\chi^2 
\nonumber \\ 
&& 
\times 
\left[ 
( \partial_z f^{(n)}_{a_{0}})^2 - \frac{3}{z^2} (f^{(n)}_{a_{0}})^2 \right]
\,, \nonumber \\ 
B_{\sigma_{gg}}^n &=& 
\frac{N_c}{12\pi^2}  \cdot \frac{1}{s^2} \cdot 
 \int \frac{d z}{z}  \left( \frac{L}{z} \right)^2 v_\chi^2 
( \partial_z f^{(n)}_{\sigma_{gg}})^2  
\,, \nonumber \\ 
C_{\sigma_{qq}\sigma_{gg}}^{(m, n)} 
&=& 
\frac{N_c}{12\pi^2} 
\int d z  \, 4 \left( \frac{L}{z} \right)^3 v_\chi^2 
 f^{(n)}_{\sigma_{gg}} 
\nonumber \\ 
&& \times 
\left[ -(\partial_z v_{qq})(\partial_z f^{(m)}_{\sigma_{qq}}) + \frac{3}{z^2} v_{qq} f^{(m)}_{\sigma_{qq}} \right] 
 \,. \label{coeffs}
\end{eqnarray} 
In deriving the effective Lagrangian we have used the equation of motion for $v_{qq}$ and $v_{\chi}$ in Eq.(\ref{EOM:v}) 
and imposed the IR boundary conditions for $f^{(n)}_{\sigma_{qq}}$, 
$f^{(n)}_{a_{0}}$ and 
$f^{(n)}_{\sigma_{gg}}$ so as to eliminate the IR boundary terms in quadratic order of fields as follows: 
\begin{eqnarray} 
&& 
\partial_z f^{(n)}_{\sigma_{qq}}(z)|_{z=z_m} = \left[ - \frac{24 \pi^2}{N_c} \lambda \xi^2 + 3- \frac{4G}{1+G} \right] f^{(n)}_{\sigma_{qq}}(z_m) 
\,, \nonumber \\ 
&& 
\partial_z f^{(n)}_{a_{0}}(z)|_{z=z_m} = \left[ - \frac{24 \pi^2}{N_c}  \lambda \xi^2 + 3- \frac{4G}{1+G} \right] f^{(n)}_{a_{0}}(z_m) 
\,, \nonumber \\ 
&& f^{(n)}_{\sigma_{gg}}(z_m) = 0 
\,, \label{IRBC:f}
\end{eqnarray} 
where use has been made of the stationary condition for $v_{qq}$ in Eq.(\ref{stationary}). 
It should be noted that the $\bar{q}q$-$gg$ mixing strength $C_{\sigma_{qq} \sigma_{gg}}^{(m, n)} $ in Eq.(\ref{coeffs}) 
is proportional to the chiral condensate $\sim v_{qq}$. 
This implies that the $\bar{q}q$-$gg$ mixing is turned off when the chiral symmetry is restored, where 
one can realize the definitely pure $\bar{q}q$ and $gg$-isospin singlet-scalar states, 
as will be discussed in the next section.

\subsection{Vector and axialvector sectors}

The five-dimensional vector and axial-vector gauge fields $V_M$ and $A_M$ are defined 
as 
\begin{equation} 
 V_M = \frac{L_M + R_M}{\sqrt{2}} 
 \,, \qquad 
A_M = \frac{L_M-R_M}{\sqrt{2}}
\,. 
\end{equation} 
It is convenient to work with the gauge-fixing $V_z=A_z\equiv 0$ and take the boundary conditions 
$V_\mu(x,\epsilon)=v_\mu(x)$, $A_\mu(x,\epsilon)=a_\mu(x)$ and  $\partial_z V_\mu(x,z)|_{z=z_m}=\partial_z A_\mu(x,z)|_{z=z_m}= 0$, 
where $v_\mu(x)$  and $a_\mu(x)$ correspond to sources for the vector and axial-vector currents, respectively. 
We then solve the equations of motion for (the transversely polarized components of) $V_\mu(x,z)$ and $A_\mu(x,z)$ 
and substitute the solutions back into the action in Eq.(\ref{S5}), to obtain the 
generating functional $W[v_\mu, a_\mu]$ holographically dual to QCD. 
Then we obtain the vector and the axial-vector current correlators, which are 
defined as 
\begin{eqnarray}
&& i \int d^4x e^{iqx} \langle 0 \vert\, {\rm T}\,  J_{V,A}^{a \mu}(x)\, J_{V,A}^{b \nu}(0)\, \vert 0 \rangle
\nonumber \\ 
&& 
= \delta^{ab} \left( \frac{q^\mu q^\nu}{q^2} - \eta^{\mu\nu} \right) \Pi_{V,A}(-q^2)  
\,, 
\end{eqnarray}
with the currents 
\begin{eqnarray}
J_V^{a\mu} &=& \bar{q} \left(\frac{T^a}{\sqrt{2}}\right) \gamma^\mu q, 
\nonumber \\
J_A^{a\mu} &=& \bar{q} \left(\frac{T^a}{\sqrt{2}}\right) \gamma^\mu \gamma_5 q.
\end{eqnarray} 
$\Pi_{V}(Q^2)$ and $\Pi_{A}(Q^2)$ (where $Q\equiv \sqrt{-q^2}$ is the Euclidean momentum)  
are expressed as 
\begin{eqnarray}
  \Pi_V(Q^2) &=& \frac{N_c}{12\pi^2} \frac{\partial_z V(Q^2, z)}{z} \Bigg|_{z=\epsilon \to 0}
  \,, \nonumber \\
    \Pi_A(Q^2) &=& \frac{N_c}{12\pi^2} \frac{\partial_z A(Q^2, z)}{z} \Bigg|_{z=\epsilon \to 0}
  \,,\label{PiVA}
\end{eqnarray} 
where the vector and axial-vector profile functions 
$V(Q^2,z)$ and $A(Q^2,z)$ are defined as $V_\mu(q,z)=v_\mu(q) V(q^2)$ and $A_\mu(q,z)=a_\mu(q) A(q^2)$ 
with the Fourier transforms of $v_\mu(x)$ and $a_\mu(x)$. 
These profile functions satisfy the following equations:   
\begin{equation} 
\left[ 
 - Q^2 + \omega^{-1} \partial_z \omega \partial_z  
\right] V(Q^2, z) = 0 ,
\label{EOM:PiV} 
\end{equation}
\begin{equation} 
\left[ 
 - Q^2 + \omega^{-1} \partial_z \omega \partial_z  - 2 \left( \frac{L}{z} \right)^2 v_{qq}^2 
\right]A(Q^2, z) = 0 ,
\label{EOM:PiA}
\end{equation}
\begin{equation} 
\omega \equiv \frac{L}{z} v_\chi^2  
\,,
\end{equation} 
with the boundary conditions 
$V(Q^2,z)|_{z=\epsilon \rightarrow 0}=A(Q^2,z)|_{z=\epsilon \rightarrow 0}=1$ 
and 
$\partial_z V(Q^2, z)|_{z=z_m}=\partial_z A(Q^2, z)|_{z=z_m}=0$.

The vector and axial-vector current correlators, $\Pi_V$ and $\Pi_A$, 
can be expanded in terms of towers of 
the vector and axial-vector resonances. 
We then identify poles for $\Pi_{V,A}$ as the infinite towers of $\rho$ and 
$a_1$ mesons. Their masses, $m_{\rho_n}$ and $m_{(a_1)_n}$, 
 are calculated by solving the eigenvalue equations for 
the vector and axial-vector profile functions~\cite{Haba:2010hu}:  
\begin{equation} 
\left[ 
 m_{\rho_n}^2 + \omega^{-1} \partial_z \omega  \partial_z  
\right] V_n(z) = 0 ,
\label{mrho:eq}
\end{equation}
\begin{equation} 
\left[ 
 m_{(a_1)_n}^2  + \omega^{-1}  \partial_z \omega  \partial_z  
- 2 \left( \frac{L}{z} \right)^2 v_{qq}^2
\right]A_n(z) = 0 ,
\label{ma1:eq}
\end{equation}
with the same boundary conditions $V_n(\epsilon)=A_n(\epsilon)=0$ and $\partial_z V_n(z)|_{z=z_m}=\partial_z A_n(z)|_{z=z_m}=0$ . 
Note that the $\rho_n$ and $(a_1)_n$ meson masses get degenerate when the chiral condensate $\sim v_{qq} \sim \xi$ is sent to zero.

The pion decay constant $f_\pi$ is expressed in terms of $\Pi_V$ and $\Pi_A$ as 
$
  f_\pi^2 = \Pi_V(0) - \Pi_A(0)  
$.  
We can express this $f_\pi$ by using Eq.(\ref{PiVA}) as~\cite{Haba:2010hu}: 
\begin{eqnarray} 
  f_\pi^2 &=&  
 - \frac{N_{c}}{12\pi^2}  
\frac{\partial_z A(Q^2,z)}{z} \Bigg|_{z=\epsilon \to 0}
 \,. \label{Fpi} 
\end{eqnarray}

The vector and axialvector sectors are completely fixed once the parameters 
$\xi$, $G$ and $z_m$ are chosen to be certain values. 
In Refs.~\cite{Haba:2010hu} and~\cite{Kurachi:2013cha}, the optimal values are found 
so as to reproduce the experimental values of $f_\pi$, 
masses of the lowest vector and axialvector mesons $m_{\rho_1} \equiv m_\rho$ and $m_{(a_1)_n} \equiv m_{a_1}$:  
the optimal numbers leading to the realistic point are 
\begin{eqnarray} 
\xi \simeq 3.1\,,\qquad 
 G \simeq 0.25 \,, \qquad 
 z_m^{-1} \simeq 347\,{\rm MeV} 
 \,. \label{opt:vals}
\end{eqnarray}
This parameter choice yields    
$f_\pi \simeq 92$ MeV, $m_\rho \simeq 775$ MeV, $m_{a_1} \simeq 1264$ MeV~\cite{Eidelman:2004wy} and predicts 
\begin{equation} 
\langle - \bar{q}q \rangle \simeq (277 \, {\rm MeV})^3 
\,, \qquad 
\langle \frac{\alpha_s}{\pi} G_{\mu\nu}^2 \rangle^{1/4} \simeq (331 \,{\rm MeV})^4
\,, \label{qq-gg-cond}
\end{equation} 
in agreement with values estimated in Refs.~\cite{Shifman:1978bx,Gasser:1982ap}~\footnote{
Our estimate on the chiral and gluon condensates is also consistent with values obtained based on 
other QCD calculations~\cite{Kataev:1981gr}
}.

\section{Realization of pure $\bar{q}q$ and $gg$ states and their mixing} 
\label{The-induced-effective-Lagrangian}

As noted above, in the present model the $\bar{q}q$-$gg$ mixing term is proportional to 
the chiral condensate $\sim v_{qq}$ [See Eq.(\ref{coeffs})]. 
When one considers the ideal limit where the chiral symmetry is restored ($\langle \bar{q}q \rangle \to 0$, $f_\pi \to 0$), 
therefore, the pure $\bar{q}q$ and $gg$-states can be unambiguously defined. 
In the present model this limit can actually be achieved 
 by sending the chiral-condensate parameter $\xi$ to zero [See Eq.(\ref{v:approx})].

\subsection{Meson mass spectra in the restoration limit of the chiral symmetry}

In the chiral-restoration limit, 
the masses of the pure $\bar{q}q$ and $gg$ isospin-singlet scalars are determined 
by the following decoupled eigenvalue equations derived from ${\cal L}_{\rm scalar}$ in Eq.(\ref{L:scalar}) with $\xi=0$:  
\begin{eqnarray} 
&& 
\left[  v_\chi^{-2} \partial_z \left( \frac{1}{z^3} v_\chi^2 \partial_z  \right) 
 + \frac{(m_{qq(n)}^0)^2}{z^3} + \frac{3}{z^5}  
\right] [f_{\sigma_{qq}}^{(n)}]^0 = 0 
\,, \nonumber \\ 
&& 
\left[  v_\chi^{-2} \partial_z \left( \frac{1}{z^3} v_\chi^2 \partial_z  \right) 
 + \frac{(m_{gg(n)}^0)^2}{z^3}  
\right] [f_{\sigma_{gg}}^{(n)}]^0  = 0 
\,, \label{eigenvalue-eq:xi0}
\end{eqnarray}
where the superscript 0 denotes $\xi=0$ ($\langle \bar{q}q \rangle = f_\pi =0$). 
  Note that, in the chiral-restoration limit the wavefunction for 
the isospin-triplet scalar $a_{0}^{(n)}$ obeys the same 
  mass eigenvalue equation as that of the isospin-singlet scalar $\sigma_{qq}^{(n)}$.     
Since the same boundary condition 
is imposed on $[f_{\sigma_{qq}}^{(n)}]^0$ and $[f_{a_{0}}^{(n)}]^0$ as in Eqs.(\ref{UVBC:f}) and (\ref{IRBC:f}), 
the isospin-triplet and -singlet quarkonia are completely degenerate in the limit $\xi\to 0$ ($\langle \bar{q}q \rangle \to 0$ and $f_\pi \to 0$).

In addition, the vector and axialvector mesons satisfy the same eigenvalue equations in the limit $\xi \to 0$ 
[See Eqs.(\ref{mrho:eq}) and (\ref{ma1:eq})], so their masses are degenerate as well. 
Note the explicit independence of $\xi$ on the eigenvalue equation for $\rho_n$ in Eq.(\ref{mrho:eq}). 
We may therefore fix the parameters $G$ and $z_m$ to the optimal values in Eq.(\ref{opt:vals}) so that 
the $\rho$ meson mass already reaches the realistic value in the chiral-restoration limit $\xi \to 0$, i.e., 
\begin{equation} 
m_\rho^0 = m_{a_1}^0 = m_\rho \simeq 775 \, {\rm MeV}
\,, 
\end{equation} 
where, again, the superscript 0 stands for $\xi=0$.

Taking $\xi=0$, $G\simeq 0.25$ and $z_m^{-1}\simeq 347$ MeV, 
we thus calculate the scalar meson masses in the chiral restoration limit. 
[One should note that the dependence of the potential parameter $\lambda$ in the IR boundary condition Eq.(\ref{IRBC:f}) goes away when $\xi=0$.]
The lowest-three scalar masses below 2 GeV are found to be 
\begin{eqnarray} 
 && 
m_{qq(1)}^0 = 0\,, 
  \nonumber \\ 
 && 
m_{qq(2)}^0 \simeq 1782 \, {\rm MeV} 
 \,, \nonumber \\ 
&&  m_{gg(1)}^0 \simeq 1782 \, {\rm MeV} 
\,.  
\end{eqnarray} 
 The lightest quarkonia, for both isospin-singlet and -triplet scalars, thus become massless in the chiral-restoration limit, 
reflecting the chiral partner of pions.

\subsection{Defining mixing between pure $\bar{q}q$ and $gg$ states}

 Once the chiral condensate develops from zero, the isospin-singlet $\bar{q}q$ and $gg$ states start to mix 
 according to the mixing form given in ${\cal L}_{\sigma_{qq} \sigma_{gg}}$ in  Eq.(\ref{Lags:scalar}). 
After the scalar fields are canonically normalized, the scalar mass terms in Eq.(\ref{L:scalar}) take the form 
\begin{eqnarray} 
&& {\cal L}_{\rm scalar}^m  
= 
\nonumber \\ 
&& - 
\sum_n \frac{1}{2} \left[ (m_{\sigma_{qq}}^{(n)})^2  (\sigma_{qq}^{(n)})^2 + (m_{a_{0}}^{(n)})^2 (a_0^{i(n)})^2 \right] 
\nonumber \\ 
&& - \sum_n \frac{1}{2} (m_{\sigma_{gg}}^{(n)})^2 (\sigma_{gg}^{(n)})^2 
- \sum_{k, n} (m_{\sigma_{qg}}^{(k,n)})^2 \sigma_{qq}^{(k)} \sigma_{gg}^{(n)}  
\,, 
\end{eqnarray}
where 
\begin{eqnarray} 
 (m_{\sigma_{qq}}^{(n)})^2 &=& 
 \frac{B_{\sigma_{qq}}^n}{A_{\sigma_{qq}}^n} 
 \,, \nonumber\\ 
  (m_{a_0}^{(n)})^2 &=& 
 \frac{B_{a_0}^n}{A_{a_0}^n} 
\,, \nonumber \\ 
 (m_{\sigma_{gg}}^{(n)})^2 &=& 
 \frac{B_{\sigma_{gg}}^n}{A_{\sigma_{gg}}^n} 
\,, \nonumber \\ 
 (m_{\sigma_{qg}}^{(k,n)})^2 &=& 
 \frac{C_{\sigma_{qq} \sigma_{gg}}^{(k,n)}}{\sqrt{A_{\sigma_{qq}}^k} \sqrt{A_{\sigma_{gg}}^n}}  
 \,.  \label{masses:holo}
\end{eqnarray} 
Note, however, that the scalar fields $\sigma_{qq}^{(n)}$ and $\sigma_{gg}^{(n)}$ are no longer 
purely $\bar{q}q$ and $gg$-states due to the nonzero mixing triggered by the nonzero $\xi$.

For the purpose of addressing the mixing between the pure $\bar{q}q$ and $gg$-states, 
we shall now propose an effective mixing term inspired by the present holographic QCD.  
We first replace the mixed wavefunctions $f_{\sigma_{qq}}^{(n)}$ and $f_{\sigma_{gg}}^{(n)}$ in 
the off-diagonal mass-squared element $(m_{\sigma_{qg}}^{(k,n)})^2$ 
of Eq.(\ref{masses:holo}) with the pure $\bar{q}q$ and $gg$ wavefunctions in the chiral-restoration limit ($\xi \to 0$),   
$[f_{\sigma_{qq}}^{(n)}]^0$ and $[f_{\sigma_{gg}}^{(n)}]^0$ in the decoupled equations in (\ref{eigenvalue-eq:xi0}). 
Then, the off-diagonal mass-squared element $(m_{\sigma_{qg}}^{(k,n)})^2$ is modified as  
\begin{eqnarray} 
(m_{\sigma_{qg}}^{(k,n)})^2 
& \to &  
 \frac{C_{\sigma_{qq} \sigma_{gg}}^{(k,n)}}{\sqrt{A_{\sigma_{qq}}^k} \sqrt{A_{\sigma_{gg}}^n}} \Bigg|_{[f_{\sigma_{qq}}^{(n)}]^0, [f_{\sigma_{gg}}^{(n)}]^0} 
\nonumber \\ 
&=& 
2 \sqrt{2} \, s \, 
\frac{\langle [f_{\sigma_{gg}}^{(n)}]^0 \left( - \dot{\bar{v}}_{qq} [f_{\sigma_{qq}}^{(k)}]^0 + \frac{3 {\bar{v}}_{qq} [f_{\sigma_{qq}}^{(k)}]^0}{z^2} \right)\rangle}
{\sqrt{ \langle [(f_{\sigma_{qq}}^{(k)}]^0)^2   \rangle \langle [(f_{\sigma_{gg}}^{(n)}]^0)^2 \rangle }} 
\,, \nonumber\\ 
\label{mixing-strength}
\end{eqnarray}    
where $\bar{v}_{qq} \equiv L \cdot v_{qq}$ and $\langle A \rangle \equiv \int dz (L/z)^3 v_\chi^2(z) A(z)$ for an arbitrary function $A(z)$. 
Thus the redefined  $(m_{\sigma_{qg}}^{(k,n)})^2 $ as in Eq.(\ref{mixing-strength}) 
is evaluated as the mixing amplitude of the pure $\bar{q}q$ state overlapped with the pure $gg$ state, 
which are properly 
defined in the chiral-restoration limit, to be developed by nonzero $\xi$ only through the nonzero $v_{qq}$.

On the other hand, we keep the diagonal elements $(m_{\sigma_{qq}}^{(n)})^2$ and $(m_{\sigma_{gg}}^{(n)})^2$ unchanged 
and assume that the wavefunctions in these elements satisfy the decoupled equations similar to Eq.(\ref{eigenvalue-eq:xi0}) 
in the chiral-restoration limit, but with the different IR boundary condition for $\sigma_{qq}^{(n)}$ having nonzero $\xi$ 
as in Eq.(\ref{IRBC:f}).

We have also computed the mass eigenvalues directly by 
solving the mixed equations, and checked that these drastic deformations do not 
make significant difference in the scalar mass spectrum.  
This implies that the mixed wavefunctions $f_{\sigma_{qq}}^{(n)}$ and $f_{\sigma_{gg}}^{(n)}$ are well saturated by 
the pure wavefunctions  
$[f_{\sigma_{qq}}^{(n)}]^0$ and $[f_{\sigma_{gg}}^{(n)}]^0$ in the scalar mass spectrum.

Even when the chiral condensate grows from zero, 
the isospin-triplet scalars $a_{0}^{i(n)}$ are still isolated as well as in the case of the chiral-restoration limit, 
except the different IR boundary condition with nonzero term $\lambda \xi^2 $ in Eq.(\ref{IRBC:f}). 
We may fix the IR potential parameter $\lambda$ so that the lowest $a_0$ mass is set to the desired value  
at the realistic point achieved by the parameter choice in Eq.(\ref{opt:vals}). 
  As done in Ref.~\cite{Haba:2010hu},  
for the reference value of the $a_0$ mass we may choose $m_{a_0} \simeq 1.24$ GeV by taking into account that  
the mass can be lifted up to that of $a_0(1450)$ when the mixing with a four-quark state is incorporated~\cite{Black:1999yz}. 
Then, the optimal value of $\lambda$ is fixed to be~\cite{Haba:2010hu}  
\begin{equation} 
 \lambda \equiv \frac{N_c}{(4\pi)^2} \cdot \kappa 
 \,, \qquad {\rm with} \qquad \kappa \simeq 1 
\,. \label{kappa:val} 
\end{equation}

\section{Analysis of the mixing structure} 
\label{Analysis-of-the-mixing-structure}

\begin{figure}[t]
\begin{center}
   \includegraphics[scale=0.50]{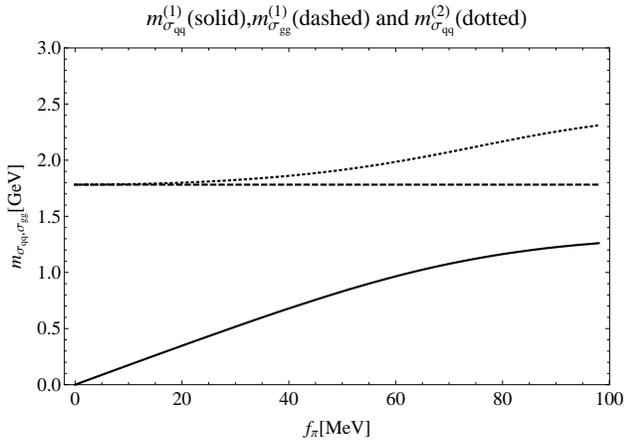}
\caption{ 
The diagonal mass-squared elements $m_{\sigma_{qq}}^{(1)}(=m_{a_0})$ (solid curve), $m_{\sigma_{qq}}^{(2)}$ (dotted curve) 
and $m_{\sigma_{gg}}^{(1)}$ (dashed curve) versus the pion decay constant $f_\pi$ 
with the parameter $G$, $z_m^{-1}$ and $\lambda$ fixed to be the optimal values, $G=0.25$, $z_m^{-1}=347$ MeV and $\lambda=3/(4\pi)^2$ 
(See Eqs.(\ref{opt:vals}) and (\ref{kappa:val})).  
\label{mqq-mgg-fpi-plot}}
\end{center} 
 \end{figure}

We are now ready to discuss the mixing strength between the pure $\bar{q}q$ and $gg$ states, 
which evolves from zero at the chiral-restoration limit ($\xi=0$) to the realistic point ($\xi \simeq 3.1$) 
with the parameter choice in Eqs.(\ref{opt:vals}) and (\ref{kappa:val}).  
In Fig.~\ref{mqq-mgg-fpi-plot} we first show the evolution of the diagonal mass-squared elements 
$(m_{\sigma_{qq}}^{(n)})^2$ and $(m_{\sigma_{gg}}^{(n)})^2$ 
with respect to $f_\pi$, in place of $\xi$. Here, we have only picked up the low-lying three mass values, 
$m_{\sigma_{qq}}^{(1)}(=m_{a_0})$, $m_{\sigma_{qq}}^{(2)}$ and $m_{\sigma_{gg}}^{(1)}$. 
The figure tells us that the lowest-$\bar{q}q$ scalar mass dramatically develops from zero at the chiral restoration point ($f_\pi=0$) 
to the fixed $m_{a_0} \simeq 1.24$ GeV at the realistic point $(f_\pi\simeq 92\,{\rm MeV})$. 
 The growth of the mass will actually be saturated around the value $\simeq$ 1.3 GeV, even if 
one extremely takes the large $f_\pi$.  
The lowest-$gg$ scalar mass $m_{\sigma_{gg}}^{(1)}$ is completely independent of $f_\pi$, while 
the mass of the first-excited $\bar{q}q$-state, $m_{\sigma_{qq}}^{(2)}$, 
goes beyond that of the ground state of the $gg$-scalar, to be above 2 GeV at the realistic point, but 
not to be above 3 GeV even when $f_\pi \to \infty$ limit.

\begin{figure}[t]
\begin{center}
   \includegraphics[scale=0.50]{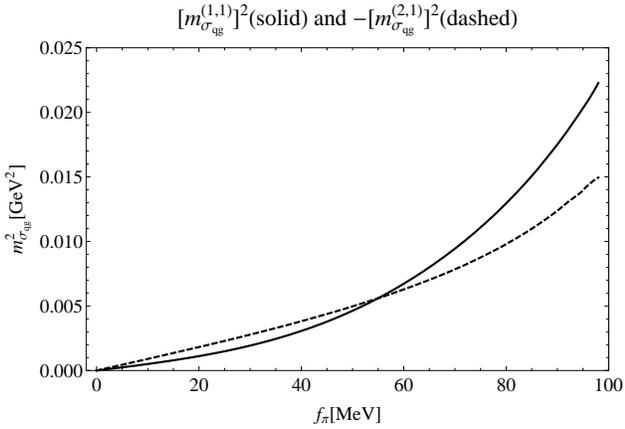}
\caption{ 
The same for the mixing strengths $(m_{\sigma_{qg}}^{(1,1)})^2$ (solid curve)
and $[ - (m_{\sigma_{qg}}^{(2,1)})^2]$ (solid curve) as Fig.~\ref{mqq-mgg-fpi-plot}. 
\label{mqg2-fpi-plot}}
\end{center} 
 \end{figure}

In Fig.~\ref{mqg2-fpi-plot} we display the mixing strengths $(m_{\sigma_{qg}}^{(1,1)})^2$ 
and $(m_{\sigma_{qg}}^{(2,1)})^2$ as a function of $f_\pi$.   
 We see from the figure that the mixing strengths are still small enough, even after reaching the realistic point, 
compared to the diagonal mass-squared values: the mass-squared matrix in unit of ${\rm GeV}^2$ looks like 
\begin{equation} 
 {\bf m}^2 \simeq \left( 
\begin{array}{ccc} 
(1.25)^2 & (0.14)^2 & 0 \\ 
(0.14)^2 & (1.77)^2 & - (0.12)^2 \\ 
0 & - (0.12)^2 & (2.26)^2  
\end{array}
\right)
\,, \label{matrix}
\end{equation}
which acts on the vector $(\sigma_{qq}^{(1)}, \sigma_{gg}^{(1)}, \sigma_{qq}^{(2)})^T$. 
By diagonalizing this matrix, one obtains the masses of the 
low-lying two isospin-singlet scalars below 2 GeV, 
\begin{equation} 
 m_{1} \simeq 1.25 \, {\rm GeV} 
\,, \qquad 
m_{2} \simeq  1.77 \, {\rm GeV} 
\,. 
\end{equation}
The lowest mass can be lifted up to the amount of the $f_0(1370)$ mass $\simeq 1.3 - 1.4$ GeV in a way similar to the $a_0(1450)$ case 
through mixing with a four-quark state~\cite{Black:1999yz}. 
Thus, compared to the diagonal element in Eq.(\ref{matrix}), we find that 
the mixing strengths are negligibly small, in accord with the assumption made in Ref.~\cite{Haba:2010hu} and 
the large $N_c$ picture on which the holographic model has been based.

The smallness of the mixing can be understood by somewhat smaller chiral-condensate scale than the gluon condensate scale: 
  these condensates are actually related involving the parameters $\xi$, $G$ and $z_m$ as~\cite{Haba:2010hu}  
  \begin{equation} 
   \xi  = \frac{32}{3 \sqrt{3}} \frac{G}{1+G} \frac{\langle - \bar{q}q \rangle/z_m}{\langle \frac{\alpha_s}{\pi} G_{\mu\nu}^2 \rangle} 
   \, . 
  \end{equation}
  Taking $G\simeq 0.25$, $\langle \frac{\alpha_s}{\pi} G_{\mu\nu}^2 \rangle \simeq (331\,{\rm MeV})^4$  
and $z_m^{-1} \simeq 347$ MeV, we numerically evaluate $\xi$ to write  
   \begin{eqnarray} 
    \xi   \simeq  5 \times \frac{\langle - \bar{q}q \rangle}{(331 \,{\rm MeV})^3} 
    \simeq  5 \times \left(  \frac{\Lambda_{\bar{q}q}}{\Lambda_{gg}} \right)^3
   \,,  \label{ratio}
  \end{eqnarray}
with the chiral and gluon condensate scales $\Lambda_{\bar{q}q}\equiv \langle - \bar{q}q \rangle^{1/3}$ 
and $\Lambda_{gg} \equiv \langle \frac{\alpha_s}{\pi} G_{\mu\nu}^2 \rangle^{1/4} $. 
 At the realistic point, from Eq.(\ref{qq-gg-cond}) we have $(\Lambda_{\bar{q}q}/\Lambda_{gg})^3 \simeq (0.84)^3 \simeq 0.6$, 
 in accordance with the optimal value of $\xi$ in Eq.(\ref{opt:vals}). 
 If $(\Lambda_{\bar{q}q}/\Lambda_{gg})^3$ was as much as/more than of ${\cal O}(1)$, $\xi$ would get larger 
 to enhance the mixing strengths to be comparable with the diagonal element in Eq.(\ref{matrix}), 
as expected from Fig.~\ref{mqg2-fpi-plot}.

 Our result thus implies that $f_0(1710)$ is predominantly constructed from the glueball state 
with the mass around $\simeq 1.7 - 1.8$ GeV, 
in agreement with the pure glueball mass estimate by lattice simulations~\cite{Morningstar:1999rf}, 
a recent study based on a holographic QCD of top-down type~\cite{Brunner:2015oqa} and  
a different approach based on a phenomenological model~\cite{Janowski:2014ppa}.

\section{Summary and discussion}
\label{Summary}

In this paper, we studied the mixing structure of isospin-singlet scalars, the light quarkonium $(\bar{q}q)$
and glueball $(gg)$ 
in two-flavor QCD, based on a holographic model of bottom-up hard-wall type. 
 Based on that model, 
we have demonstrated that the $\bar{qq}$-$gg$ mixing takes place due to the nonzero chiral condensate. 
In the model, the pure quarkonium and glueball states are unambiguously defined in terms of the different $U(1)_A$ charges 
in the restoration limit of the chiral $U(2)_L \times U(2)_R$ symmetry, in which the quarkonium gets massless as the chiral partner of the pion.  
We showed that, 
at the realistic point where the pion decay constant and other hadron masses reach the observed amount, 
the lightest quarkonia and glueball are hardly mixed at all,      
without any phenomenological inputs such as the currently observed isospin-singlet scalar-decay properties. 
This is 
consistent with the large $N_c$ picture on that of basis the holographic model has been established. 
The smallness of the mixing strength can actually be understood by the slightly smaller ratio of the chiral and gluon condensates (See Eq.(\ref{ratio})). 
The low-lying two scalar masses are calculated to be $\simeq 1.25$ GeV and $\simeq 1.77$ GeV, 
which are compared with masses of $f_0(1370)$ and $f_0(1710)$. 
Our result implies that $f_0(1710)$ predominantly consists of glueball.

Several comments are in order:

In addressing the scalar meson masses, 
we have incorporated not only the lowest masses from each of $\bar{q}q$ and $gg$ states, but also the next-to-lowest one 
from the $\bar{q}q$ state. 
As seen from the mass matrix element in Eq.(\ref{matrix}), the mixing between the two lightest scalars from the $\bar{q}q$ and $gg$ 
could be comparable with that between the lightest scalar from the $gg$ and the next-to-lightest one from the $\bar{q}q$. 
 This would imply some new possibility that incorporating the next-to-lowest $\bar{q}q$ scalar could be relevant 
in investigating the low-lying scalar mixing structure.

Also, we have neglected the higher level more than the third in the quarkonium state  
and the second in the glueball state. Actually, we have computed the higher masses to be $\gtrsim 3$ GeV, which has not  
yet been established in experiments on the scalar resonances, but could be accessible in the future, together with 
the slightly lighter $\bar{q}q$-like scalar with the mass $\simeq 2$ GeV predicted in the present model (See Eq.(\ref{matrix})).

 Similar analyses as done in the present paper can be performed 
in a more realistic situation, where mixings with a four-quark state and 
 a pure strange state with the sizable strange quark mass are incorporated by extending to the three-flavor case, 
which will be pursued in another publication.

\acknowledgments

We are grateful to M.~Harada and H.~Nishihara for useful comments and discussions. 
S.M was supported in by the JSPS 
Grant-in-Aid for Scientific Research (S) No. 22224003.

\end{document}